\documentclass{article}
\usepackage{ace,amsmath,graphicx,url,times}
\usepackage{pgfplots}
\pdfoutput=1

\title{Estimation of the direct-to-reverberant  Energy Ratio using a spherical microphone array}

\name{Hanchi Chen, Prasanga N. Samarasinghe, Thushara D. Abhayapala, Wen Zhang}
\address{Research School of Engineering\\
       College of Engineering and Computer Science\\
       The Australian National University}

\begin{document}

\ninept
\maketitle

\begin{sloppy}

\begin{abstract}
This paper proposes a practical approach to estimate the direct-to-reverberant energy ratio (DRR) using a spherical microphone array without having knowledge of the source signal. We base our estimation on  a theoretical relationship between the DRR and the coherence estimation function between coincident pressure and particle velocity. We discuss the proposed method's ability to estimate the DRR in a wide variety of room sizes, reverberation times and source receiver distances  with appropriate examples. Test results show that the method can estimate the room DRR for frequencies between $199$  - $2511$ Hz, with $\pm 3$ dB accuracy.
\end{abstract}

\section{Introduction}
\label{sec:intro}

The direct-to-reverberant energy ratio (DRR) is a  helpful tool to characterize reverberant enclosures. The DRR is also  important in speech processing and amplification applications, such as hearing aids, speech recognition, or teleconferencing, as it either influences the algorithm strategy or its success. Further, in psychoacoustics the DRR is believed to be an important cue for the perception of source-to-receiver distance.

Traditionally, the DRR is  directly calculated from the room impulse response.  This approach is less feasible in practical scenarios as it requires measurement of the impulse response. For this reason, several authors have recently investigated the use of alternative blind-methods based on statistical approaches. Larsen \textit{et al.} \cite{larsen2003acoustic} and Falk \textit{et al.} \cite{falk2010temporal} have proposed alternate methods  to calculate the DRR based on estimated room impulse responses. Even though these methods perform well, they both require a priori processing,  and obtaining the respective prior information might be too difficult due to the unstable nature of acoustic environments. Another approach for DRR estimation based on a binaural system was proposed in \cite{lu2010binaural} which uses directional filtering to extract the direct path. The main drawback of this method is ignoring the fact that there could be reverberant sound coming from the direct path itself. A statistical approach based on spatial coherence matrix of a microphone array output was recently introduced in \cite{hioka2011estimating}, where the broadband signal power of direct and reverberant components are estimated using a least-squared method. This requires an estimation of the direct path's direction of arrival (DOA) and therefore requires a priori processing, that is crucial to the DRR estimation.

To avoid pre-processing altogether, Kuster \cite{Kuster_01} later proposed a purely analytical approach based on a mathematical relationship between the DRR and and the magnitude-squared coherence estimation function between coincident pressure and particle velocity. Even though the DOA appears in the above relationship, it was shown that reliable results can be obtained when it's set to $0^{\circ}$ as long as the true angle between the direct path and velocity component is smaller than $60^{\circ}$. The microphone array used consisted of two omnidirectional sensors and the particle velocity was estimated from the gradient between them.

In this paper, we adopt the  theory developed by Kuster, but use a spherical microphone array consisting of $32$ omnidirectional sensors (Eigenmike).  To apply the particle velocity concept in the spherical harmonic/modal domain, we first derive  three first order harmonics, which  individually represent velocities along three orthogonal directions at the origin \cite{evers2014multiple}. These are then used to derive  the desired velocity component along the direct path. Instead of setting the DOA to $0^{\circ}$ as in \cite{Kuster_01}, we estimated it utilizing the frequency smoothed MUSIC algorithm for maximum accuracy.

\section{Preliminary: sound field recording using Eigenmike}
 The Eigenmike consists of 32 condenser microphones mounted on the surface of a rigid sphere. Therefore we can use the rigid sphere model to express the sound pressure on the surface of the Eigenmike.

 The sound pressure at a point on the surface of a rigid sphere is the combination of two components: the impinging wave and the scattered wave. If we define a spherical coordinate with its origin located at the center of a sphere with radius $R$, using the spherical harmonic decomposition, the sound pressure $P(r,\theta,\phi,k)$ at a point on the surface of the sphere can be expressed as \cite{FourierAcoustics, ward2001performance}
 \begin{equation}                                                                     \label{harm1}
  P(R,\theta,\phi,k)=\\
  \sum_{n=0}^\infty \sum_{m=-n}^N \alpha_{nm}(k)b_n(kR)Y_{nm}(\theta,\phi),
 \end{equation}
 with
 \begin{equation}
 b_n(kR)=j_n(kR)-\frac{j'_n(kR)}{{h_n^{(2)}}'(kR)}h_n^{(2)}(kR),
 \end{equation}
 where $k=2\pi f/c$ is the wave number, $f$ and $c$ are the frequency and the wave propagation speed, respectively. $\alpha_{nm}$ are the spherical harmonic coefficients, $j_n(kR)$ is the spherical Bessel function of order $n$, $h_n^{(2)}(kR)$ is the spherical Hankel function of the second kind with order $n$, and $Y_{nm}(\theta,\phi)$ denotes the spherical harmonic of order $n$ and degree $m$. $Y_{nm}(\theta,\phi)$ has the orthogonal property
 \begin{equation}                                                                      \label{eq:orthogonal}
 \int_0^\pi\int_0^{2\pi} Y_{nm}(\theta,\phi)Y_{n'm'}^*(\theta,\phi)\sin{\theta} d\theta d\phi=\delta_{n-n',m-m'}.
 \end{equation}
 In the case of the Eigenmike, $R=42$mm. Due to the orthogonal property \eqref{eq:orthogonal}, the spherical harmonic coefficients may be calculated by
 \begin{equation}                                                                             \label{eq:int coef}
 \alpha_{nm}(k)=\int \frac{P(R,\theta,\phi,k)Y^*_{nm}(\theta,\phi)}{b_n(kR)}\sin{\theta}d\theta d\phi.
 \end{equation}
 The discrete version of \eqref{eq:int coef} may be used to calculate the coefficients using the sound pressure information measured by the 32 microphones on the Eigenmike,
 \begin{equation}                                                                             \label{eq:sum coef}
 \alpha_{nm}(k)=\sum_{i=1}^{32} W_i \frac{P(R,\theta,\phi,k)Y^*_{nm}(\theta,\phi)}{b_n(kR)}\sin{\theta}d\theta d\phi
 \end{equation}
 where $W_i$ are weighing factors corresponding to each microphone.

\section{Direction of Arrival estimation}
In order to estimate the direction of arrival (DOA) of the direct path, we used the MUSIC algorithm in the spherical harmonic domain \cite{Abhayapala_01, khaykin2009coherent}. The main advantage of processing in the spherical harmonic/modal  domain, is the decoupling of frequency-dependent and angular-dependent components. We  exploited this result to de-correlate any coherent signals by performing frequency smoothing (averaging) over the broadband spectrum \cite{Abhayapala_01, khaykin2009coherent, abhayapala_11}. This last step is considered to be of high importance as the coherence between direct and reverberant signals are high. Based on the rigid array configuration discussed in Section~\ref{sec:intro}, the MUSIC spectrum was defined by
\begin{equation}
\label{eq:MUSICspectrum}
f_{\text{MUSIC}}(\theta,\phi)=\frac{1}{\boldsymbol{y}(\theta,\phi)\boldsymbol{E}_{n}\boldsymbol{E}_{n}^H\boldsymbol{y}^H(\theta,\phi)}
\end{equation}
\noindent where $\boldsymbol{y}(\theta,\phi)= [Y'_{00}(\theta,\phi) \cdots \cdots Y'_{NN}(\theta,\phi)]^H$ is the steering vector of the Eigenmike and $\boldsymbol{E}_{n}$ is a matrix containing the noise eigenvectors of the frequency smoothed modal cross spectrum
\begin{equation}
\label{eq:modalCS}
\boldsymbol{R}=\frac{1}{I}\sum_{i=1}^{I}\boldsymbol{\alpha_{i}}\boldsymbol{\alpha^{H}_{i}}
\end{equation}

\noindent with  $\boldsymbol{\alpha_{i}}=[\alpha_{00}(k_{i}) \cdots \alpha_{NN}(k_{i})]$ representing the soundfield coefficients (\ref{eq:sum coef}) at frequency $k_{i}$ ($i=1 \cdots I$). As MUSIC assumes the noise and signal subspaces to be orthogonal, the maximum peak of (\ref{eq:MUSICspectrum}) was considered as the DOA estimate of the direct path speech signal.

\section{DRR estimation using particle velocity and sound pressure measurements}
 A method of estimating DRR using particle velocity measurements in addition to sound pressure measurements has been proposed in \cite{Kuster_01}. In this section, we briefly outline this algorithm.

 Assuming that both the direct and reveberant sound are plane waves, the total particle velocity and sound pressure at a point can be expressed as \cite{Kuster_01}
 \begin{equation}
 V(k)=\frac{A_D}{\rho_0c}e^{j\varphi_0}\cos{\vartheta_0}+\sum_{i=1}^{\infty}\frac{A_R}{\rho_0c}e^{j\varphi_i}\cos{\vartheta_i},
 \end{equation}
 \begin{equation}
 P(k)=A_D e^{j\varphi_0}+\sum_{i=1}^{\infty}A_Re^{j\varphi_i},
 \end{equation}
 where $A_D$ and $A_R$ are the magnitudes of the direct path and reverberant paths, respectively; $\vartheta_0$ and $\vartheta_i$ are the angles between the particle velocity and the direct / reverberant path impinging directions, respectively; $\varphi_0$ and $\varphi_i$ are the phases of each impinging wave. Assuming that the reverberation is evenly distributed for all directions, with random phases, the following assumptions can be made:
 \begin{align}
 & E\{e^{j\varphi_0}\sum_{i=1}^\infty e^{j\varphi_i}\}=0, \\
 & E\{\sum_{i=1}^\infty \cos{\vartheta_i}\}=0, \\
 & \sum_{i=1}^\infty |A_R|^2=|A_D|^2/\text{DRR},
 \end{align}
 where $E\{\cdot\}$ is the mathematical expectation.

 The spectral densities can then be expressed as
 \begin{align}
 &S_{PP}=E\{PP^*\}=|A_D|^2(1+\frac{1}{\text{DRR}}),\\                                \label{eq:density1}
 &S_{PV}=E\{PV^*\}=\frac{|A_D|^2}{\rho_0c}\cos{\vartheta_0},\\
 &S_{VV}=E\{VV^*\}=\frac{|A_D|^2}{\rho_0^2c^2}(\cos{\vartheta_0}^2+0.5/{\text{DRR}}).
 \end{align}
 Finally, the spatial coherence function between the particle velocity and sound pressure can be expressed using \eqref{eq:density1},
 \begin{align}                     \label{eq:gamma}
 \gamma & =\frac{|S_{PV}|^2}{S_{PP}\times S_{VV}}\\
 & = \frac{(\text{DRR}\cos{\vartheta_0})^2}{(1+\text{DRR})(0.5+\cos{\vartheta_0}^2)}.
 \end{align}
 Solving for $\text{DRR}$ yields
 \begin{multline}                                         \label{eq:DRR solution}
 \text{DRR}= -\frac{\gamma^2+2\cos{\vartheta_0}^2\gamma^2}{4\cos{\vartheta_0}^2(\gamma^2-1)}\\
-\frac{2\gamma\sqrt{\cos{\vartheta_0}^4\gamma^2-\cos{\vartheta_0}^2\gamma^2+2\cos{\vartheta_0}^2+0.25\gamma^2}}{4\cos{\vartheta_0}^2(\gamma^2-1)}.
 \end{multline}
 Given the measured data of sound pressure and particle velocity, the value of $\gamma$ may be calculated using \eqref{eq:gamma}. Substituting the calculated $\gamma$ into \eqref{eq:DRR solution} yields the final DRR estimation.
 %
\section{Estimation of particle velocity using spherical harmonic coefficients}
%
 Typically, the particle velocity of the sound wave is measured by a microphone with differential beam pattern. In this work, we use the Eigenmike to synthesis the beam pattern of a differential microphone, so as to evaluate the particle velocity along an arbitrary direction.

 The beam pattern of a differential microphone is a bi-polar pattern which can be expressed as
 \begin{equation}
 G(\vartheta_d)=\cos{\vartheta_d}
 \end{equation}
 %
 where $\vartheta_d$ is the angle between the direction of microphone principal axis $(\theta_{\text{mic}},\phi_{\text{mic}})$ and the impinging direction of the sound wave $(\theta,\phi)$. Suppose that the differential microphone is placed such that its principal axis points at $(\theta_{\text{mic}}=\pi/2,\phi_{\text{mic}}=0)$, or in other words parallel to the $x$ axis in the Cartesian coordinate system, then the beam pattern can be expressed using spherical harmonics,
 \begin{equation}
 G_x(\theta,\phi)=\frac{Y_{1 1}(\theta,\phi)+Y_{1 (-1)}(\theta,\phi)}{2}
 \end{equation}
 Similarly, the same beam pattern with its principal axis parallel to $y$ axis $(\theta_{\text{mic}}=\pi/2,\phi_{\text{mic}}=\pi/2)$ and $z$ axis $(\theta_{\text{mic}}=0,\phi_{\text{mic}}=0)$ can be expressed as
 \begin{equation}
 G_y(\theta,\phi)=\frac{Y_{1 1}(\theta,\phi)-Y_{1 (-1)}(\theta,\phi)}{2i}
 \end{equation}
 and
 \begin{equation}
 G_z(\theta,\phi)=Y_{1 0}(\theta,\phi),
 \end{equation}
 respectively. Using these beam weights, the particle velocity components of any impinging wave at these three directions can be represented using the spherical harmonic coefficients:
 \begin{align}
 & V_x=(\alpha_{1 1}+\alpha_{1 (-1)})/2, \\
 & V_y=(\alpha_{1 1}-\alpha_{1 (-1)})/2i, \\
 & V_z=\alpha_{1 0}.
 \end{align}
 %
 Using these velocity components, we can synthesis the particle velocity measured at an arbitrary direction $(\theta_v,\phi_v)$ by summing the projection of these velocity vectors on the desired direction,
 %
 \begin{equation}                                                                       \label{eq:velocity}
 V(\theta_v,\phi_v)=V_x\sin{\theta_v}\cos{\phi_v}+V_y\sin{\theta_v}\sin{\phi_v}+V_z\cos{\theta_v}.
 \end{equation}
 %
 In addition, it can be seen from \eqref{harm1} that by setting $R=0$, the sound pressure at the origin is equal to the $0$th order coefficient, i.e., $P_o=\alpha_{0 0}$.
 %
\subsection{Spatial averaging of the DRR estimation}
%
 The assumptions used to derive \eqref{eq:DRR solution} are valid if the reverberant sound field is isotropic; however, in reality this condition may not be fulfilled, which results in an error in the DRR estimation. In addition, since we use the Eigenmike to synthesis the particle velocity measurement, it is possible to simultaneously synthesis the velocity measurement at multiple directions. In our experiments, we found out that by estimating the DRR using particle velocity measurements in multiple directions simultaneously and taking their average, the final result is much more accurate and consistent over different frequency bands.

 Suppose the direct path impinging angle from the DOA estimation is $(\theta_d,\phi_d)$, then, using \eqref{eq:velocity}, we calculate the particle velocities in multiple different directions. Then, we calculate the DRR for each velocity direction using \eqref{eq:DRR solution}, with $\vartheta_0$ being the angle between the direct path and the particle velocity direction. Here, the DRR value calculated using \eqref{eq:DRR solution} is in linear scale, therefore the final log-scale DRR after spatial averaging is calculated as
 \begin{equation}
 \text{DRR}=10\log_{10}\big( \frac{\sum_{i=1}^K \text{DRR}_i}{K}\big).
 \end{equation}
 where $\text{DRR}_i$ represents the linear scale DRR estmation based on velocity measurement at the $i$th direction, and $K$ is the total number of estimations being calculated.

 The effect of spatial averaging is illustrated in FIG.~\ref{fig:Average}. The test data file ``EM32\_Purple\_A\_Bob\_work\_Fan\_20dB.wav'' from the ACE database \cite{Eaton_01} is used to generate this figure. The figure depicts the subband DRR estimation results acquired using a single velocity measurement (green curve), pointed at the direct path direction $(\theta_d=145^\circ, \phi_d=90^\circ)$, as well as the DRR estimation using the spatial averaging method (red curve), with four velocity measurements pointed at $(\theta_d+60^\circ,\phi_d)$, $(\theta_d-60^\circ,\phi_d)$, $(\theta_d,\phi_d+60^\circ)$ and $(\theta_d,\phi_d-60^\circ)$. We also include the ground truth (blue curve) for comparison.

 \begin{figure}[t]
  \centering
  \centerline{\includegraphics[width=\columnwidth]{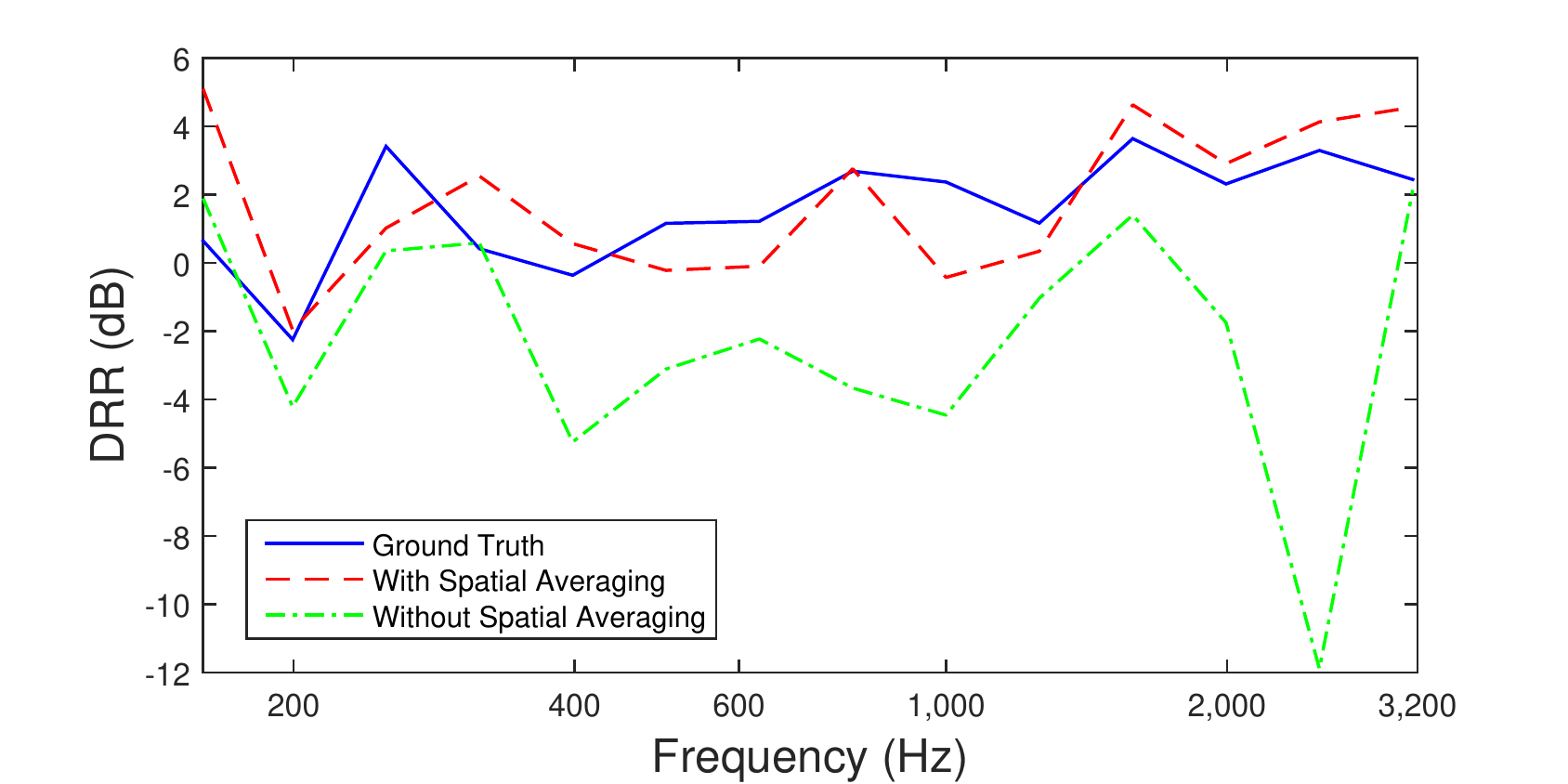}}
  \caption{Comparison of DRR estimation accuracy with and without spatial averaging.}
  \label{fig:Average}
 \end{figure}

 It can be seen from FIG.~\ref{fig:Average} that the spatial averaging method effectively improves the DRR estimation accuracy, this is particularly true for the higher frequency bands. The DRR estimation using single velocity measurement is $~4$ dB lower than the ground truth in the $400$-$800$ Hz band, and a significant drop can be observed at $2500$ Hz. In comparison, the spatial averaged estimation show less than $2$ dB deviation from the ground truth in most frequency bands. Our experiments show that adding more velocity estimations show minimum improvement compared to the setup using 4 estimations, therefore the setup used in FIG.~\ref{fig:Average} is used for all our DRR calculations.

\section{Performance Analysis}
\subsection{Subband DRR Estimation}
 We use the proposed algorithm to estimate the DRR at multiple frequency bands, the central frequencies are between $199$ Hz and $2511$ Hz. This range of frequency roughly covers the spectrum of human voice. For lower frequencies, our algorithm is unable to reliably estimate the DRR due to lack of speech signal energy. Also, we derive the particle velocity measurements from the spherical harmonic coefficients, and at very low frequencies, the Eigenmike cannot acquire the first order harmonics with high accuracy, while at very high frequencies, spatial aliasing prevents accurate acquisition of any spherical harmonic coefficient.

 We use the proposed method to estimate the subband DRR for each recording in the ACE Challenge database.  The mean value and the standard deviation of the estimation errors for recordings with $18$ dB SNR are shown in FIG.~\ref{fig:Sub18}.
 \begin{figure}[t]
  \centering
  \centerline{\includegraphics[width=\columnwidth]{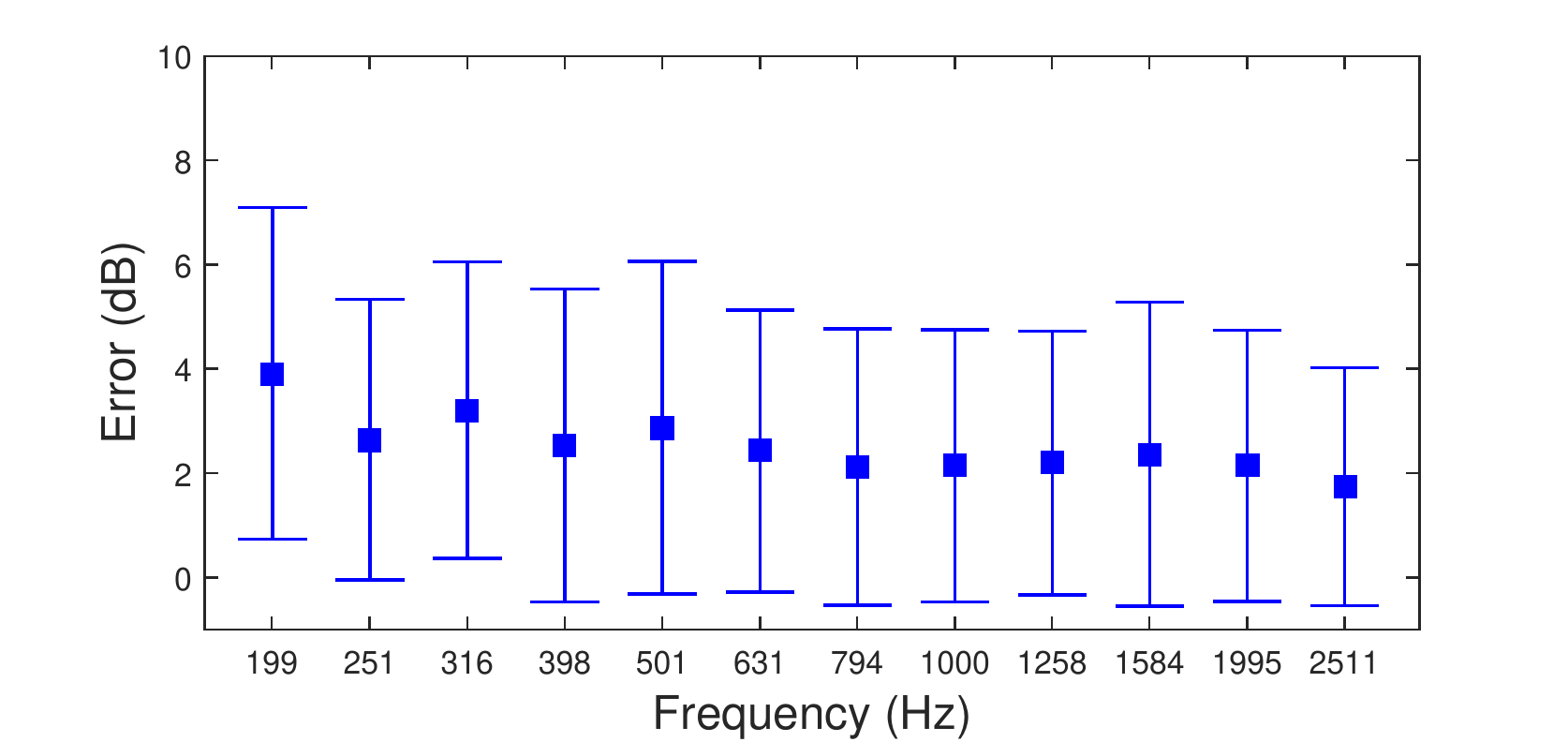}}
  \caption{Subband DRR estimation error for all rooms and configurations at $18$ dB SNR.}
  \label{fig:Sub18}
 \end{figure}

 It can be seen from FIG.~\ref{fig:Sub18} that nearly all the estimated DRR mean values fall within $\pm3$ dB of the ground truth, and the standard deviation for each band is below $3$ dB. Furthermore, the subband error shows a decreasing trend as the frequency increases.

 The subband results for recordings with $-1$ dB SNR are shown in FIG.~\ref{fig:Sub1}.

  \begin{figure}[t]
  \centering
  \centerline{\includegraphics[width=\columnwidth]{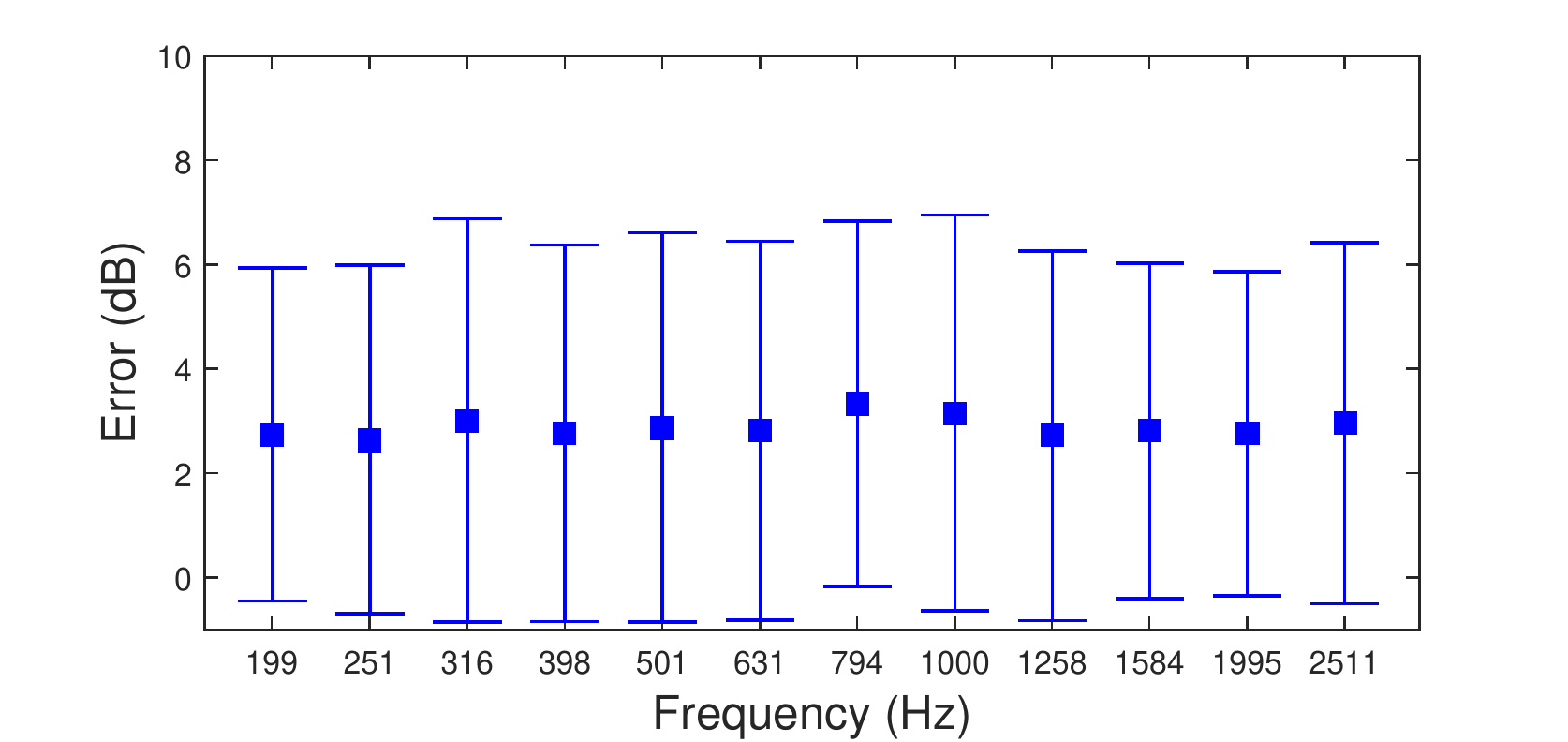}}
  \caption{Subband DRR estimation error for all rooms and configurations at $-1$ dB SNR.}
  \label{fig:Sub1}
 \end{figure}

 Comparing FIG.~\ref{fig:Sub1} with FIG.~\ref{fig:Sub18}, we can see that with increased noise level, the mean estimation error has increased for nearly all subbands. Also increased is the standard deviation of the estimation error, which is around $4$ dB for all subbands. Furthermore, the estimation accuracy no longer improves at higher frequencies, contrary to the $18$ dB SNR results. In general, we may conclude that the proposed method can estimate subband DRR with $3$ dB accuracy for both low and high SNR, although for higher SNR, the algorithm produces more accurate results especially at higher frequencies.

\subsection{Fullband DRR Estimation}
 Since our algorithm is unable to reliably estimate the DRR at very low and high frequencies, we calculate the full band DRR estimation by averaging the subband results from $199$ Hz to $2511$ Hz only. We carry out the calculation for all five rooms and two distances for each room as given in the ACE database, and compare our results with the ground truth for each case. The average error and standard deviation of our estimations for $18$ dB SNR are shown in FIG.~\ref{fig:18dB}.

\begin{figure}[t]
  \centering
  \centerline{\includegraphics[width=\columnwidth]{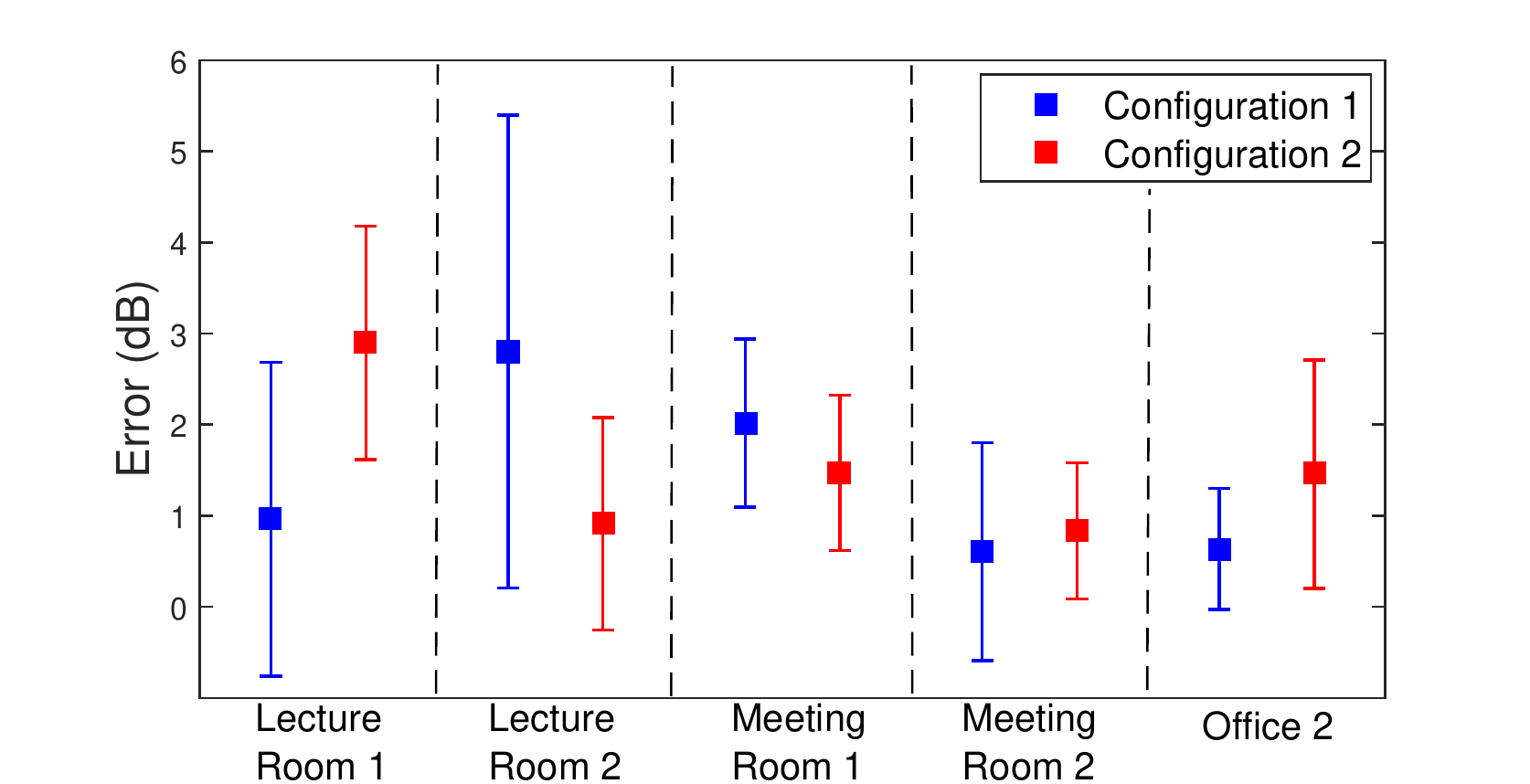}}
  \caption{Full band DRR estimation result for all rooms and configurations with $18$dB SNR.}
  \label{fig:18dB}
 \end{figure}

 We can see from FIG.~\ref{fig:18dB} that the full band estimation error for all room setups are within $3$ dB. In particular, 5 room configurations have $~1$ dB mean error, and another 3 room configurations have less than $2$ dB of error. Only two setups have a mean error of $3$ dB. The error standard deviation is also less than $2$ dB for nearly all room setups. We notice that the two room setups with the highest error standard deviation correspond to the two lecture room recordings with long source-to-microphone distances, which lead to low DRR ground truths. This indicates that the proposed algorithm may be less accurate when the reverberation is very strong.

\section{Conclusion}
 In this paper, we propose an algorithm to perform Direct-to-Reveberant Ratio estimation using an Eigenmike. Using the spherical harmonic coefficients calculated from the Eigenmike recordings, we estimate the Direction-of-Arrival of the direct path, as well as synthesis the sound pressure and particle velocity in multiple directions at the center of the Eigenmike. Finally, we use an algorithm based on coherence between particle velocities and sound pressure to estimate the DRR.

 We use the ACE evaluation database to assess the performance of our algrithm, the results show that our algorithm can reliably estimate the room DRR from approximately $199$ Hz to $2511$ Hz.

%

\end{sloppy}
\end{document}